\documentclass[twocolumn]{aastex631}
\usepackage{amsmath}

\usepackage{tabularx}
\usepackage{booktabs}

\usepackage{verbatim}

\graphicspath{{./}{}}
     
\usepackage{color}   
\makeatletter
\newcommand*{\chk}{\@ifnextchar\bgroup{\chk@}{\color{blue}}}
\newcommand*{\chk@}[1]{{\textcolor{blue}{#1}}}
\makeatother

\usepackage[utf8]{inputenc}

\def\arcmin{\hbox{$^\prime$}}
\def\arcsec{\hbox{$^{\prime\prime}$}}

\begin{document}

\title{ASKAP observations of the radio shell in the composite supernova remnant G310.6$-$1.6}

\author[0009-0004-4090-9304]{Wenhui Jing}
\affiliation{School of Physics and Astronomy, Yunnan University, Kunming, 650091, P. R. China; wh.jing@outlook.com, xhsun@ynu.edu.cn, xhli@ynu.edu.cn}

\author[0000-0001-7722-8458]{Jennifer L. West}
\affiliation{Dominion Radio Astrophysical Observatory, Herzberg Astronomy \& Astrophysics, National Research Council Canada, P.O. Box 248, Penticton, BC V2A 6J9, Canada; jennifer.west@nrc-cnrc.gc.ca}

\author[0000-0002-3464-5128]{Xiaohui Sun}
\affiliation{School of Physics and Astronomy, Yunnan University, Kunming, 650091, P. R. China; wh.jing@outlook.com, xhsun@ynu.edu.cn, xhli@ynu.edu.cn}

\author[0000-0002-6155-9962]{Wasim Raja}
\affiliation{Australia Telescope National Facility, CSIRO, Space and Astronomy, PO Box 76, Epping, NSW 1710, Australia; Wasim.Raja@csiro.au}

\author[0000-0022-9399-8433]{Xianghua Li}
\affiliation{School of Physics and Astronomy, Yunnan University, Kunming, 650091, P. R. China; wh.jing@outlook.com, xhsun@ynu.edu.cn, xhli@ynu.edu.cn}

\author[0009-0006-1753-7623]{Lingxiao Dang}
\affiliation{School of Astronomy \& Space Science, Nanjing University, 163 Xianlin Avenue, Nanjing 210023, People’s Republic of China}

\author[0000-0002-5683-822X]{Ping Zhou}
\affiliation{School of Astronomy \& Space Science, Nanjing University, 163 Xianlin Avenue, Nanjing 210023, People’s Republic of China}
\affiliation{Key Laboratory of Modern Astronomy and Astrophysics, Nanjing University, Ministry of Education, Nanjing 210023, People’s Republic of China}

\author[0000-0002-4990-9288]{Miroslav D. Filipovi\'c}
\affiliation{Western Sydney University, Locked Bag 1797, Penrith South DC, NSW 2751, Australia}

\author[0000-0002-6097-2747]{Andrew M. Hopkins}
\affiliation{School of Mathematical and Physical Sciences, 12 Wally's Walk, Macquarie University, NSW 2109, Australia}

\author[0000-0001-5953-0100]{Roland Kothes}
\affiliation{Dominion Radio Astrophysical Observatory, Herzberg Astronomy \& Astrophysics, National Research Council Canada, P.O. Box 248, Penticton, BC V2A 6J9, Canada; jennifer.west@nrc-cnrc.gc.ca}

\author[0000-0001-6109-8548]{Sanja Lazarevi\'c}
\affiliation{Western Sydney University, Locked Bag 1797, Penrith South DC, NSW 2751, Australia}
\affiliation{Australia Telescope National Facility, CSIRO, Space and Astronomy, PO Box 76, Epping, NSW 1710, Australia; Wasim.Raja@csiro.au}
\affiliation{Astronomical Observatory, Volgina 7, 11060 Belgrade, Serbia}

\author[0000-0002-4814-958X]{Denis Leahy}
\affiliation{Department of Physics and Astronomy University of Calgary, Calgary, AB, T2N 1N4, Canada}

\author[0000-0002-9994-1593]{Emil Lenc}
\affiliation{Australia Telescope National Facility, CSIRO, Space and Astronomy, PO Box 76, Epping, NSW 1710, Australia; Wasim.Raja@csiro.au}

\author[0000-0003-0742-2006]{Yik Ki Ma}
\affiliation{Research School of Astronomy \& Astrophysics, The Australian National University, Canberra, ACT 2611, Australia}
\affiliation{Max-Planck-Institut f\"ur Radioastronomie, Auf dem H\"ugel 69, 53121 Bonn, Germany}

\author[0000-0002-7641-9946]{Cameron L. Van Eck}
\affiliation{Research School of Astronomy \& Astrophysics, The Australian National University, Canberra, ACT 2611, Australia}

\begin{abstract}

We report the observations of the radio shell of the supernova remnant (SNR) G310.6$-$1.6 
at 943 MHz from the Evolutionary Map of the Universe (EMU) and the Polarization Sky Survey of the Universe's Magnetism (POSSUM) surveys by using the Australian Square Kilometre Array Pathfinder (ASKAP). 
We detect polarized emission from the central pulsar wind nebula (PWN) with rotation measures varying from $-$696~rad~m$^{-2}$ to $-$601~rad~m$^{-2}$. We measure the integrated flux density of the shell to be $36.4\pm2.2$~mJy at 943~MHz and derive a spectral index of $\rm \alpha_{pwn} = -0.4\pm0.1$ for the PWN and $\rm \alpha_\text{shell} = -0.7\pm0.3$ for the SNR shell. From the combined radio and X-ray observations, the object can be identified as a supernova explosion of about 2500~yr ago with energy of about $1.3\times10^{50}$~erg, suggesting an ejected mass of about $10\,M_\sun$. The circular radio shell outside the circular hard X-ray shell is unique among Galactic SNRs. 
We discuss several possible scenarios, including blast wave, reverse shock, and pulsar-fed emission, but find that none of them can fully explain the observed characteristics of the shell. This poses a challenge for understanding the evolution of SNRs. 
The results of this paper demonstrate the potential of the ASKAP EMU and POSSUM surveys in discovering more objects of small angular size and low surface brightness. 

\end{abstract}

\section{Introduction} \label{sec:introduction}

Composite supernova remnants (SNRs) consist of a shell and a pulsar wind nebula (PWN) powered by relativistic particles from the associated pulsar~\citep{Dubner2015,Vink2020}.  
According to the catalog compiled by~\citet{Ferrand2012}, 
only a handful of them exhibit well-defined circular shells surrounding the PWN in radio or X-ray images. Although spherical symmetry has been suggested as a possible feature of some composite SNRs, there is no evidence to show a correlation between this symmetry and other SNR parameters, such as spectral indices, SNR ages, molecular cloud (MC) associations, or Galactic locations~\citep{Ranasinghe2023}. Each of them has shown its individual properties and has provided us with valuable insight into the SNR system.

G310.6$-$1.6 (RA=$\rm{14^h 00^m 45^s}$, Dec=$-63\arcdeg 26\arcmin$, J2000) is considered to be a composite SNR with a bright central PWN and a circular shell. It was first identified as a PWN from Chandra X-ray observations by \citet{Tomsick2009}. \citet{Renaud2010} detected the associated pulsar PSR J1400$-$6345 from both X-ray and radio timing observations, with a period of 31.18~ms. 

The characteristic age is about 12.7~kyr, and the spin-down luminosity is about $5.1\times10^{37}$~erg~s$^{-1}$ 
for the pulsar \citep{Renaud2010}, making it one of the most energetic Galactic pulsars known. \citet{Renaud2010} also discovered the faint shell from the Chandra X-ray image and thus identified G310.6$-$1.6 as a composite SNR. 
They examined archival radio observations and obtained a spectral index $\rm \alpha_{pwn} = -0.3$, defined by $S_\nu \propto \nu^\alpha$, where $S_\nu$ is the flux density at frequency $\nu$. It confirms the identification of the central object as a PWN, which is consistent with expectations \citep{Reynolds2012}.

\cite{Reynolds2019} analyzed the Chandra X-ray observations of a much longer exposure time than that in \citet{Renaud2010}, and found that the X-ray spectrum of the shell is featureless and can be well fitted with a model of synchrotron radiation. This placed G310.6$-$1.6 into the category of X-ray synchrotron SNRs (XSSNRs). XSSNRs refer to SNRs with X-ray emission from the shell dominated by synchrotron radiation~\citep[see][for a review]{Reynolds2008}.
Among the eight XSSNRs identified so far~\citep{Reynolds2024}, G310.6$-$1.6 is the only one that has also been classified as a composite SNR.

Given that the synchrotron X-ray emission is probably from the forward shock, radio emission from the shell is expected. There have been no published radio observations dedicated to G310.6$-$1.6. It was included in several previous surveys, such as the Molonglo Galactic plane survey at 843~MHz~\citep{Murphy2007}, the Parkes survey of the southern Galactic plane at 2.4~GHz~\citep{Duncan1995}, and the Parkes-MIT-NRAO survey at 4.85~GHz~\citep{Griffith1993,Condon1993}. However, limited by resolution and sensitivity, these radio images are dominated by the bright PWN, and the putative shell remains elusive. Based on these observations, an upper limit for the surface brightness of the shell was estimated to be about 2.3$\times$10$^{-21}$~W~m$^{-2}$~Hz$^{-1}$~sr$^{-1}$ at 1~GHz~\citep{Renaud2010}. We note that \citet{Robbins2014thesis} reported the detection of the radio shell in the PhD thesis based on observations with the Australian Telescope Compact Array at 5.5 and 9~GHz, but did not perform detailed analysis due to the data quality.

The Australian Square Kilometre Array Pathfinder~\citep[ASKAP,][]{Hotan2021} with its ongoing continuum survey~\citep[the Evolutionary Map of the Universe, EMU, ][Hopkins et al. in prep. ]{Norris2011, Norris2021} and polarization survey~\citep[the POlarization Sky Survey of the Universe's Magnetism, POSSUM,][Gaensler et al. 2025 submitted]{Gaensler2010} delivers images with a high sensitivity of 25-30 $\mu$Jy beam$^{-1}$ and a high resolution of about $15\arcsec$, allowing us to detect low surface-brightness emission from small, faint shells. This has been demonstrated by \citet{Ball2023}, \citet{Filipovic+2023}, and \citet{Lazarevic+2024a} who discovered new SNRs from the ASKAP data. 

In this paper, we report the observation of the radio synchrotron shell of SNR G310.6$-$1.6 with ASKAP. Data acquisition and reprocessing are described in Sect.~\ref{sec:acquisition}, the results of images and spectra are presented in Sect.~\ref{sec:results}, discussions of evolution scenarios are given in Sect.~\ref{sec:discussion}, and conclusions are drawn in Sect.~\ref{sec:conclusions}. 

\section{Data acquisition and reprocessing}\label{sec:acquisition}

\subsection{ASKAP data}

The 30-square degree field containing G310.6$-$1.6 was observed on the 30th of September, 2023 for 10 hours (ASKAP scheduling block ID 53310) using 36 electronically formed beams arranged in a hexagonal close-pack configuration in the {\it continuum averaged} mode where the total frequency bandwidth is divided into 288 1-MHz channels \citep[see Fig. 20 in][]{Hotan2021}. The central frequency of these observations is 943~MHz. The calibrated visibilities for each beam, the {\it mosaicked} multi-frequency synthesized (MFS) Stokes $I$ image, as well as the {\it mosaicked} image cubes at 1~MHz resolution for all four Stokes parameters ($I$, $Q$, $U$ and $V$) are available to public on CSIRO's ASKAP Science Data Archive (CASDA)\footnote{\url{https://data.csiro.au/collections/domain/casdaObservation/search/}}.  
The images archived on CASDA are convolved to have a matching resolution across all 36 beams. This leads to poorer than achievable resolution in parts of the mosaic. The common resolution for the archived MFS images for this observation is 15$\arcsec$, while for the data cubes the resolution is frequency-dependent with values as coarse as 18$\arcsec$ at the lowest frequencies. 
 
\subsection{Images}
\begin{figure*}[htb]
    \centering
    \includegraphics[width=0.95\textwidth]{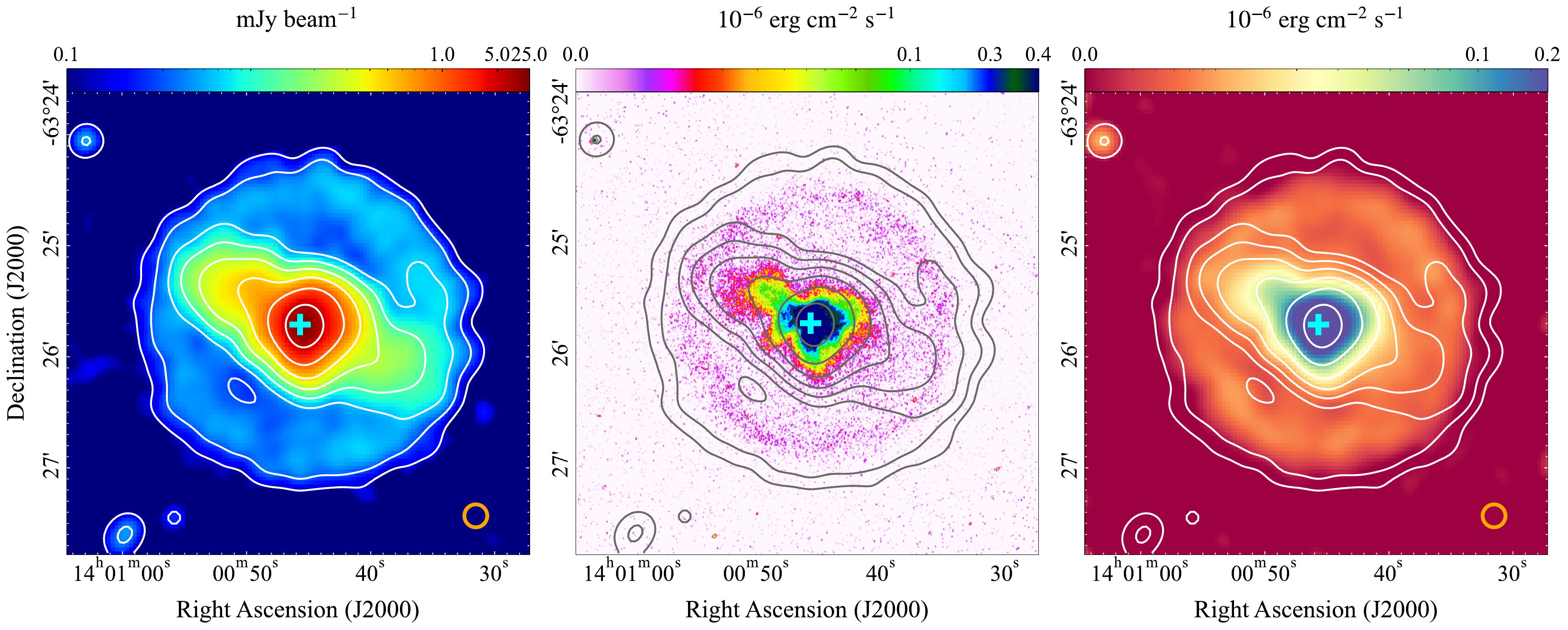}
    \caption{Left: ASKAP MFS Stokes $I$ image at 943~MHz with an angular resolution of $12\farcs4$, shown by the orange circle in the bottom right-hand corner; Middle: Chandra X-ray photon image averaged across the 1.0-8.0~keV band with an angular resolution of $0\farcs5$; 
    Right: the same as the middle panel, but with the X-ray image smoothed to $12\farcs4$. All contours show the ASKAP Stokes $I$ in the levels of $5\sigma\times 2^i,\,i=0,\,\ldots,\,5$ with the rms noise $\sigma$ of 30~$\mu$Jy~beam$^{-1}$. 
    The cyan cross indicates the position of PSR J1400$-$6325.}
    \label{fig:intensity}
\end{figure*}

{\it \textit{Full-band MFS Stokes $I$ image.}} We reprocessed the archived calibrated visibilities using ASKAPsoft~\citep{Guzman2019}. MFS images with the full 288-MHz band for Stokes $I$ from beams 27, 28 and 33, containing the SNR in their field-of-view, are generated, primary-beam corrected, and mosaicked together. The resulting Stokes $I$ image has a slightly better resolution of $12\farcs4 \times 12\farcs4$. We used this image for our analyses in the paper. 

{\it \textit{Sub-band MFS Stokes $I$ images for spectral index measurement.}} The MFS imaging algorithm generates images of the Taylor term coefficients. The images of the first two coefficients can be used to compute the spectral indices. But the derived spectral indices using the entire band could be prone to errors, especially for the weak diffused emission from the SNR \citep[for details of these issues, see Chapter 6 in the thesis][]{rauThesis}. To compute the in-band spectral indices reliably, we sub-divided the visibilities into 4 spectral windows, each 72~MHz wide, and performed MFS imaging using two Taylor terms producing Stokes $I$ images at 835, 908, 980, and 1051~MHz. These images are used in our derivation and analysis of spectral indices. 

{\it \textit{RM-Synthesis.}}: We convolved the archived Stokes $I(\nu)$, $Q(\nu)$ and $U(\nu)$ spectral data cubes available on CASDA to a common $18\arcsec$ resolution at all frequencies. We then used the RM-Tools package~\citep{Purcell2020} to perform rotation measure (RM) synthesis~\citep{Brentjens2005} and RM clean \citep{Heald+2009} to generate the $Q(\phi)$ and $U(\phi)$ data cubes at Faraday depths ($\phi$) ranging from $-2000$~rad~m$^{-2}$ to $+2000~$rad~m$^{-2}$ in steps of $5~$rad~m$^{-2}$. We then searched for peaks in $\sqrt{Q(\phi)^2 + U(\phi)^2}$ along the Faraday depth axis. A detailed description of the procedure can be found in \citet{Vanderwoude+2024}.

\subsection{Chandra data}
This SNR has been observed multiple times by the Chandra X-ray observatory since 2006. These observations were carried out with the Advanced CCD Imaging Spectrometer (ACIS) in the Very Faint mode to minimize particle background interference. 
We used all of the archived data (Chandra ObsID: 9058, 12567, 17905, 19919, 19920, and 21689) to generate the image of photon counts and conduct spectral analysis. 
Compared to the dataset of \citet{Reynolds2019}, we include two more observations (ObsIDs 9058 and 21689), which contribute an additional exposure time of $\sim$36 ks.

We analyzed the data with the Chandra Interactive Analysis of Observations (CIAO) software version 4.15 \citep{Fruscione2006}. The energy bands in the range of 1.0-8.0~keV were combined to produce the exposure-corrected and energy-filtered image. The total effective exposure time is $\sim$231.53 ks. 

\section{Results} \label{sec:results}

The reprocessed ASKAP MFS Stokes $I$ image of G310.6$-$1.6 is shown in Fig.~\ref{fig:intensity} (left panel). The frequency is 943~MHz and the resolution is $12\farcs4$. The rms noise is 30~$\mu$Jy~beam$^{-1}$. The Chandra X-ray image in the 1.0-8.0~keV bands at the original resolution of $0\farcs5$ and smoothed to the same resolution as the ASKAP image is shown in the middle and right panels, respectively, in Fig.~\ref{fig:intensity}. 

The radio emission of G310.6$-$1.6 clearly consists of two components: the central PWN and the faint radio SNR shell. The radio PWN running from southwest to northeast is aligned with the X-ray PWN and extends to the edge of the radio emission. Toward the southwest, the brightness falls more rapidly in X-rays than in radio \cite[similar to what is seen in MSH 11-62, ][]{Slane2012}.  Similarly to X-ray, the radio shell is faint and slightly limb-brightened, and has a well-defined circular shape. We derive the flux density and profile of the shell, excluding the portions that overlap with the central PWN. 

\begin{figure}[ht!]
\centering
\includegraphics[width=0.48\textwidth]
{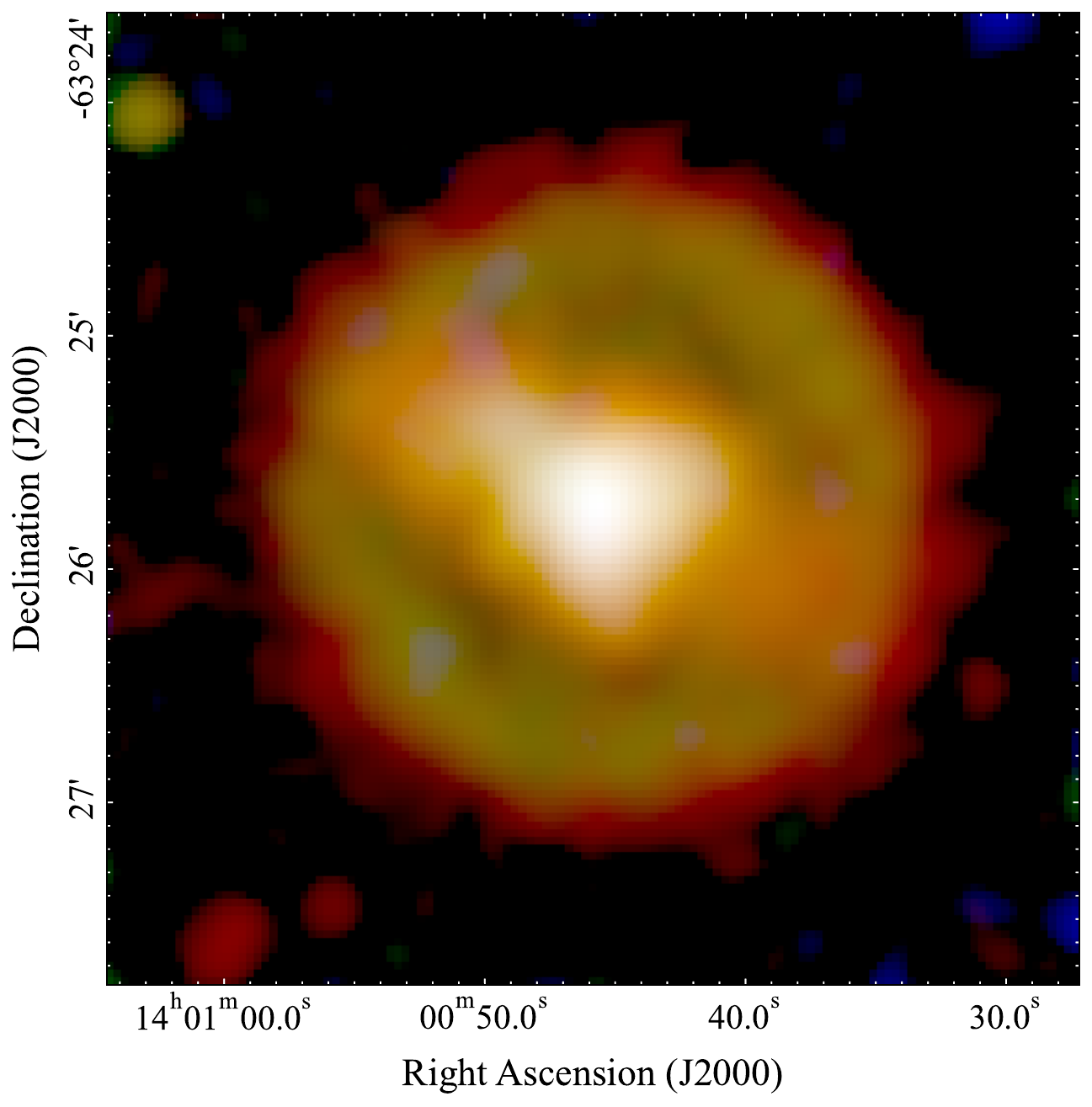}
\caption{An RGB three-color image with red representing radio emission observed with ASKAP, blue for X-ray emission observed with Chandra in the 0.5$-$1.2 keV range, and green for X-ray emission in the 1.2$-$8.0 keV range. All images have a resolution of $12\farcs4$.}
\label{fig:RC}
\end{figure}

In Fig.~\ref{fig:intensity}, we can clearly see that radio emission extends beyond X-ray emission. The mismatch between radio and X-ray emission from the shell is even more outstanding, as illustrated by the RGB three-color image in Fig.~\ref{fig:RC}, which was produced with \texttt{multicolorfits} \citep{2019ascl.soft09002C}. Here we split the Chandra band into two ranges: 0.5$-$1.2 keV (soft) and 1.2$-$8.0 keV (hard). The radio and X-ray emissions from the central PWN align well overall, but the radio emission is more extended.

\begin{figure}[ht!]
\centering
\includegraphics[width=0.49\textwidth]{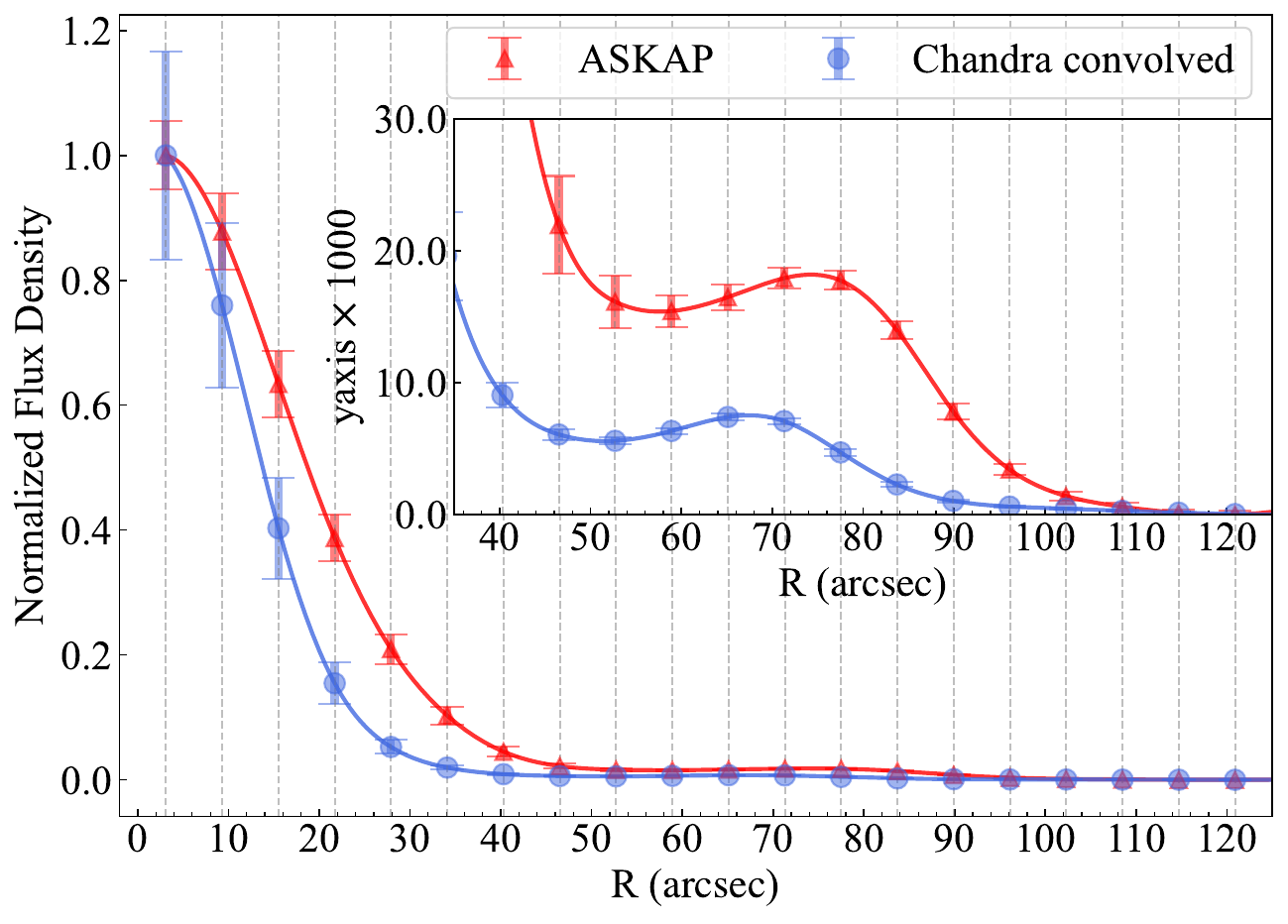}  
\caption{Radial profiles (normalized by their respective maxima) of radio intensity (red) and X-ray photon counts (blue). The inset panel is a zoom-in of the $y$-axis to show the slightly limb-brightened shell. The profiles are derived from the images with a common resolution of $12.4\arcsec$.}
\label{fig:profile}
\end{figure}

Using the same resolution images, we obtain the radial profiles of radio intensity and X-ray photon counts by averaging the values within annuli of $6\farcs2$ width from the geometric center at (RA=$\rm{14^h 00^m 45.19^s}$, Dec=$-63\arcdeg 25\arcmin 40.9\arcsec$, J2000) to a radius of $124\arcsec$. In order to circumvent the influence of the PWN to the SNR shell, we exclude the area with azimuthal angle in the ranges of $18\arcdeg$-$100\arcdeg$ and $207\arcdeg$-$273\arcdeg$, with $0\arcdeg$ defined as north and angles measured counterclockwise.  Note that both the radio intensity image and the X-ray photon counts image have a resolution of 12.4\arcsec, as presented in Fig. \ref{fig:intensity}~(left and right panels). To facilitate comparison, the profiles are normalized to a range from 0 to 1 by their respective maxima, as shown in Fig.~\ref{fig:profile}. The zoomed-in panel highlights the slightly limb-brightened shells. The radio shell extends from a radius of about $54\arcsec$ to a radius of about $100\arcsec$ with a peak at about $75\arcsec$. The X-ray shell extends from a radius of about $50\arcsec$ to a radius of approximately $90\arcsec$ with a peak at about $66\arcsec$. The ratio of the width of the shell to the radius of the shell is about 0.45, similar in both radio and X-ray emission. However, the radio shell is shifted outward compared to the X-ray shell, as can be seen from Figs.~\ref{fig:intensity} and \ref{fig:RC}.

We measure the integrated flux densities of the whole SNR, the SNR shell, and the PWN of G310$-$1.6 from the full-band and the four sub-band MFS Stokes $I$ images~(Table~\ref{tab:fluxDensity}). The integrated flux density of the shell is 36.4~mJy at 943~MHz, corresponding to a surface brightness of about $7\times 10^{-22}$~W~Hz$^{-1}$~sr$^{-1}$~m$^{-2}$, well below the upper limit estimated by \citet{Renaud2010}, explaining its non-detection in previous radio surveys.

\begin{deluxetable}{ccccc}\label{fluxDensity}
\tablecolumns{5}
\tablecaption{Integrated flux densities for the whole SNR, the SNR shell, and the PWN of G310.6$-$1.6. The uncertainties (at the 68\% confidence level) are given in parentheses. 
}
\label{tab:fluxDensity}
\tablehead{
    \colhead{Frequency} & \colhead{Bandwidth} &\colhead{Whole SNR} & \colhead{SNR shell} & \colhead{PWN} \\
    \colhead{(MHz)} & \colhead{(MHz)}& \colhead{(mJy)} & \colhead{(mJy)} & \colhead{(mJy)}
}
\startdata
 943  & 288 & 284.7 (1.2) & 36.4 (2.2) & 248.3 (1.8) \\
 835  & 72 & 298.2 (2.9) & 38.5 (2.3) & 257.8 (1.7) \\
 908  & 72 & 288.7 (2.2) & 36.6 (1.7) & 250.0 (1.3) \\
 980  & 72 & 278.2 (1.7) & 34.5 (1.3) & 242.2 (0.9) \\
 1051 & 72 & 266.3 (1.7) & 32.4 (1.1) & 232.4 (0.9) \\
\enddata
\tablecomments{The region of whole SNR and PWN are defined as the contour levels of 5$\sigma$ and 20$\sigma$ shown in Fig.~\ref{fig:intensity} (left panel). For the integrated flux density of the SNR shell, the region overlapped with PWN was excluded. }
\end{deluxetable}

\subsection{Radio spectra}\label{sec:radiospectra}
\begin{figure}[ht!]
    \plotone{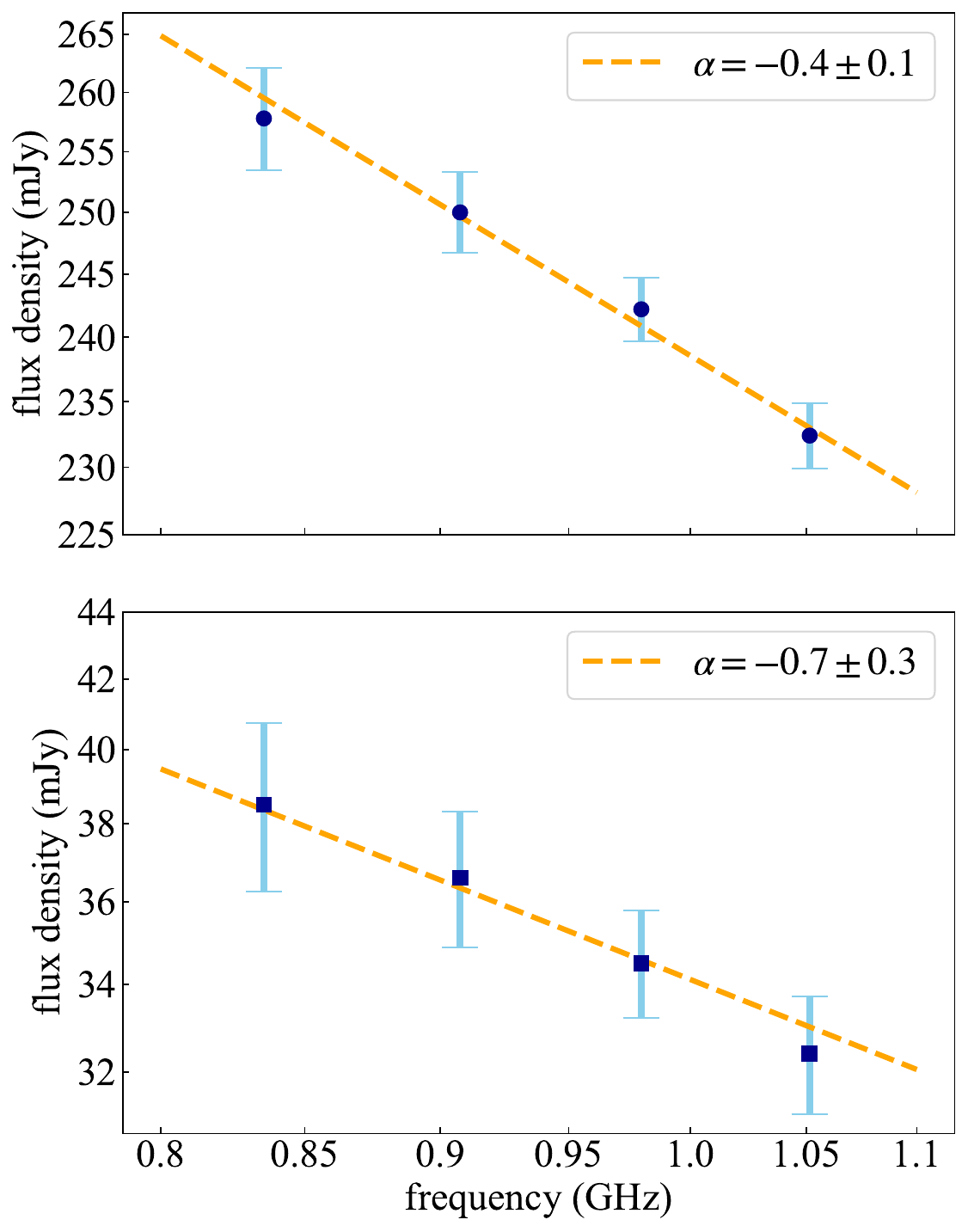}
    \caption{Radio in-band spectra for the PWN (top) and the shell (bottom) of G310.6$-$1.6 using the flux density values listed in Table~\ref{tab:fluxDensity}.
    }
    \label{fig:spec2d}
\end{figure}

The in-band spectra for the central PWN and the shell are derived using the integrated flux densities in Table~\ref{tab:fluxDensity}, and are shown in Fig.~\ref{fig:spec2d}.

We obtained a spectral index of $\rm \alpha_\text{pwn} = -0.4 \pm 0.1$ for the PWN and a spectral index of $\rm \alpha_\text{shell} = -0.7 \pm 0.3$ for the SNR shell, which fall in the range for PWNe and SNR shells, respectively \citep{Reynolds2012,Kothes2017}. The uncertainties for the spectral indices are derived using the bootstrap method. We created 1000 realizations based on the mean and error for the four data points, and performed linear fittings to obtain 1000 spectral indices. These spectral indices conform to a Gaussian distribution and its standard deviation was used as the uncertainty shown in Fig.~\ref{fig:spec2d}.

\citet{Renaud2010} fit the flux densities measured from previous observations at 843~MHz and 4.85~GHz and obtained a spectral index of $\rm \alpha_\text{pwn} = -0.33\pm0.05$ or $\rm \alpha_\text{pwn} = -0.205\pm0.008$ depending on the methods used to measure the flux density at 4.85~GHz. Their flux densities contain the contribution from the shell, but the contribution is small, only about 12\% at frequencies around 943~MHz~(Table~\ref{tab:fluxDensity}), and expected to be even smaller at higher frequencies with the steep spectrum. Therefore, their spectral index mainly represents the PWN, which is consistent with our result.  

\subsection{X-ray spectra of the shell}\label{sec:xrayspectra}
The X-ray spectral analysis by~\citet{Reynolds2019} uses the upper limit of the radio flux density at 1~GHz for the shell. Since we now have a measurement of the flux density, we use this new measurement in a reanalysis of the X-ray spectral fit.

\begin{figure}	\plotone{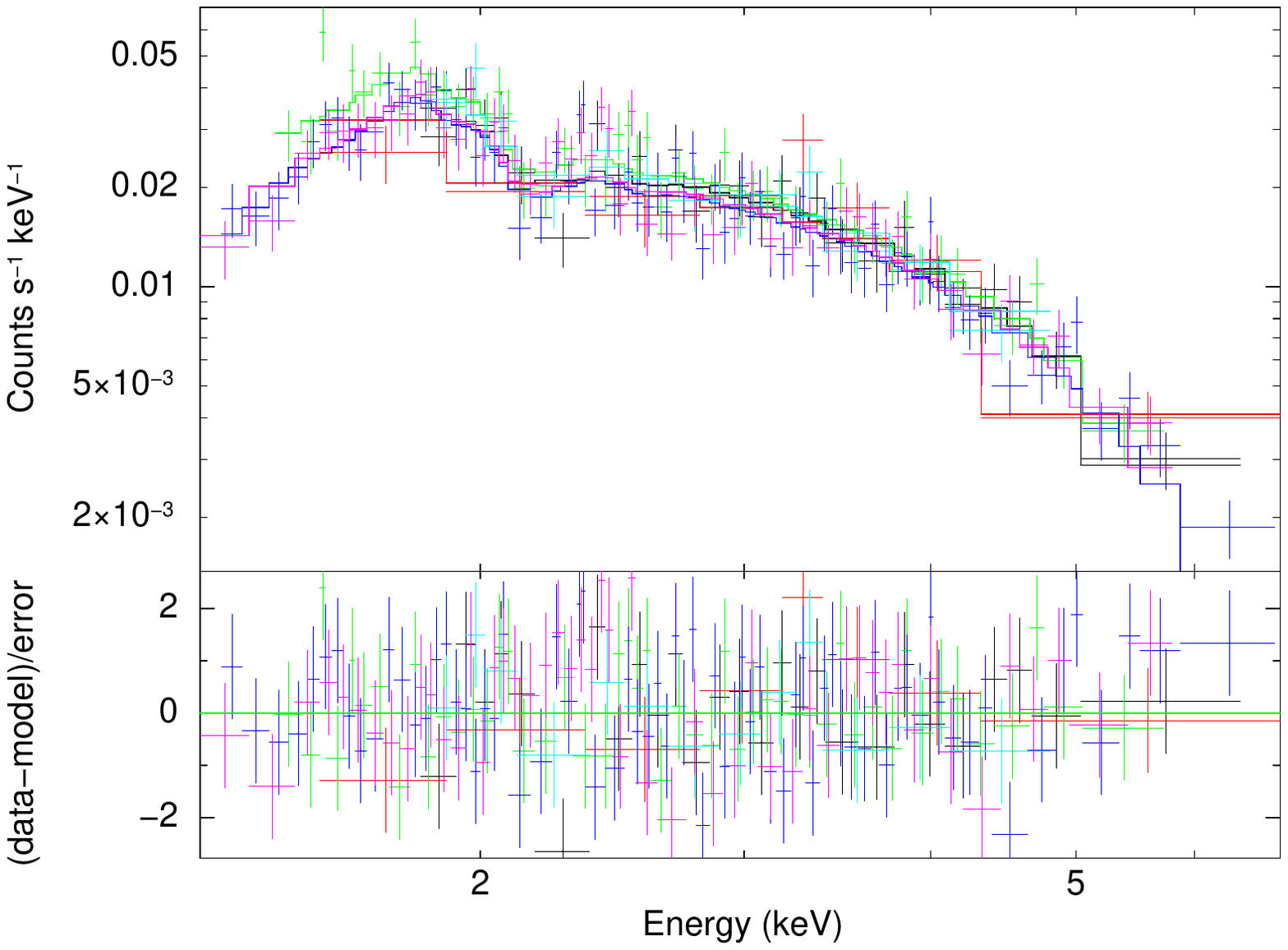}
    \caption{Shell X-ray spectra of G310.6$-$1.6 fitted with a synchrotron model \textit{srcut}. The red, green, blue, cyan, magenta and black colors correspond to the Chandra observation IDs 9058, 12567, 17905, 19919, 19920 and 21689.}
    \label{fig:srcut}
\end{figure}

We retrieve the shell spectra~(Fig.~\ref{fig:srcut}) from 6 Chandra X-ray observations (ObsIDs: 9058, 12567, 17905, 19919, 19920, and 21689) and rebin them with a minimum signal-to-noise ratio of 5. Different background regions are selected for the observations to ensure that the background spectra and the source spectra are extracted from the same CCD. 

Using XSPEC Version 12.14.0~\citep{Arnaud1996}, we jointly fit the spectra with an absorbed (tbabs) synchrotron (srcut) model with a constant component added to account for calibration uncertainty between observations (\textless10\%).
This synchrotron model describes the spectrum of electrons with an exponential cutoff power-law distribution in a homogeneous magnetic field. Note that we select the SNR shell region that clearly emits both radio and X-ray emissions, excluding areas such as the northwest and southeast parts between the PWN and the shell structure. As a result, the selected region is slightly different from that used for deriving the integrated flux density shown in Table \ref{tab:fluxDensity}. We measure the integrated flux density for this region and scale this value to 1~GHz using our measurement of $\alpha_\text{shell} = -0.7$ to obtain a value of 0.032$\pm$0.003~Jy for the model parameter ``norm''. The solar abundances of \cite{2009aspl} were adopted in the absorption model. 
As shown in Fig.~\ref{fig:srcut}, the \textit{srcut} model can fit well the shell spectra, providing a reduced $\chi^2/\rm{d.o.f.}$ of 0.97. The results are shown in Table \ref{tab:srcut}, where $\rm N_{H}$ represents the equivalent hydrogen column density and $\rm \nu_{rolloff}$ indicates the rolloff frequency, and $\rm \alpha_\text{shell}$ is the fitted radio spectral index. We note that within the uncertainties, this fitted value is in agreement with our measured value.
Varying the norm value between 29--35 mJy causes a minimal change in the fitting results. This means that the spectral index between X-ray and radio emission is much flatter than measured between radio bands. Fixing $\alpha_\text{shell}=-0.6$ or $-0.7$ results in a poor fit to the X-ray spectra. 

To assess the contribution of thermal emission, we fit the spectra with a \textit{powerlaw} plus a plane-parallel shocked plasma model (\textit{pshock}) following \cite{Reynolds2019}. Adding this thermal component does not significantly improve the spectral fit according to the F-test\footnote{https://heasarc.gsfc.nasa.gov/xanadu/xspec/manual/node82.html} (probability = 0.88). The parameters of $pshock$ model cannot be constrained, and there is a degeneracy between the electron temperature $kT_e$ and the ``norm'' value of this model ($\propto n_e^2$, with $n_e$ the electron density). We estimate the preshock density $n_0=n_e/4.8<26/0.93/0.16/0.08 f^{-0.5}$~cm$^{-3}$ for $kT_e=0.1/0.3/1/3$~keV, where $f\leq1$ is the volume filling factor of the X-ray-emitting gas. The density is different from that was obtained by \citet{Reynolds2019}. Our study adds two new Chandra spectra and uses the updated solar abundance table \citep{2009aspl}, which can cause some differences in the spectral fit, especially considering that the thermal component is not significantly detected.

\begin{deluxetable}{ccc}
\tablecolumns{2}

\tablecaption{Results from the spectral fits of shell using \textit{srcut} model.}
\label{tab:srcut}
\tablehead{
\colhead{Parameter} &  Previous study\tablenotemark{a} &\colhead{This paper}
}
\startdata
norm(mJy, input) & $ \leq 40 $ & $32$\\
$\rm N_{H} (\rm{10^{22}cm^{-2}})$ & $2.75^{+0.13}_{-0.12} $ & $3.34^{+0.21}_{-0.33}$  \\
$\rm \alpha_\text{shell}$ & $-0.49^{+0.1}_{-0.2}$&$-0.46^{+0.03}_{-0.02}$ \\
$\rm \nu_{rolloff} (Hz) $ & $(1.4^{+0.4}_{-0.7})\times 10^{17} $&  $(9.7^{+7.3}_{-2.4}) \times 10^{16}$ \\
$\chi^2/\rm{d.o.f.}$ & -- & 208.77/215  \\
\enddata
\tablenotetext{a}{The results are from \citet{Reynolds2019}}
\end{deluxetable}

\begin{figure*}[!htb]
    \includegraphics[width=0.95\textwidth]{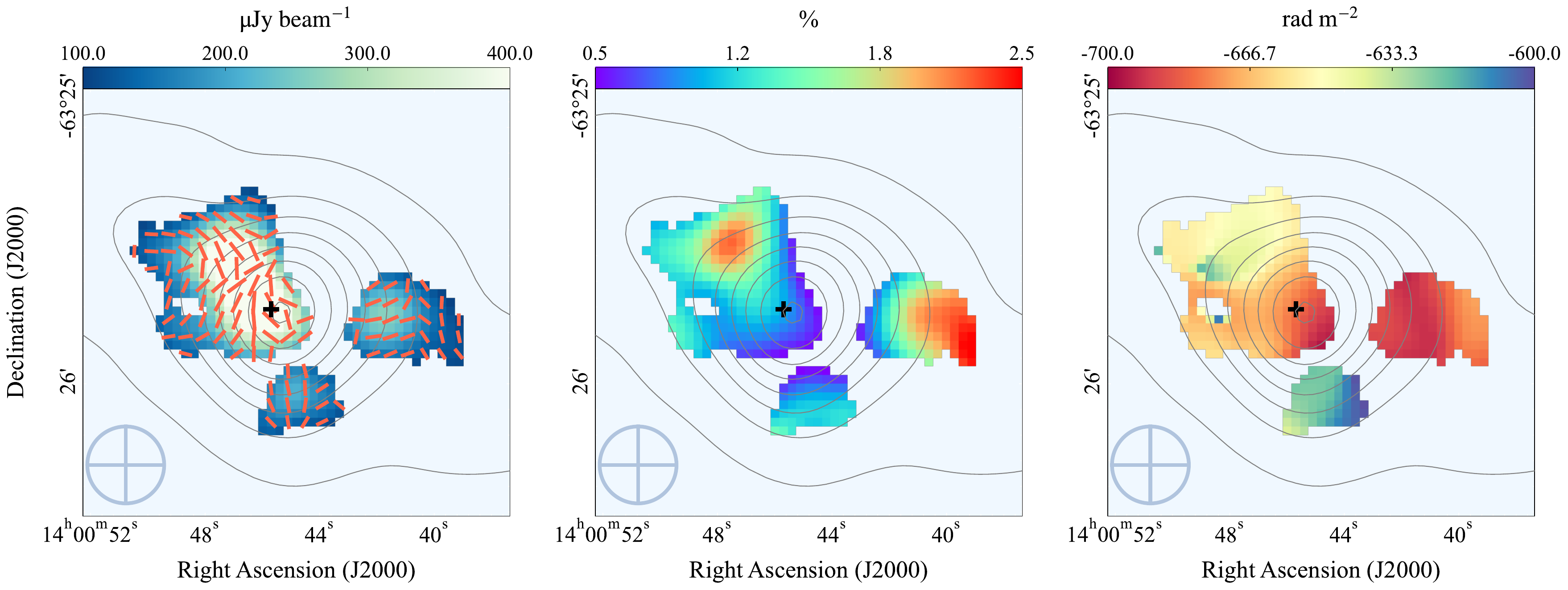}
    \centering
    \caption{Left panel: polarized intensity image of the PWN overlaid with bars indicating orientation of magnetic field corrected for RM. 
    The length of the bars has been scaled to the square root of the polarized intensity to enhance the visibility of the orientation. Middle panel: fractional polarization image. 
    Right panel: RM image. The position of PSR J1400$-$6325 is marked with a cross. The contours display the total intensity with levels starting from 2~mJy~beam$^{-1}$ and in steps of 6~mJy~beam$^{-1}$. All the images have a resolution of $18\arcsec$ shown by the blue ellipse in the bottom left-hand corner.}
    \label{fig:PI}
\end{figure*}

\subsection{Polarization}

The fitted peak polarized intensity ($PI$), the fractional polarization ($p$, defined as $p= PI/I$), and the fitted peak Faraday depth images derived from RM synthesis\footnote{Using the \url{rmtools_peakfitcube} tool from RM-Tools} are shown in Fig.~\ref{fig:PI}. In the case of G310.6$-$1.6, the strongest peak for a given sight line is the dominant emission component, and the fitted peak Faraday depth is regarded as RM. The orientation of the magnetic field corrected for Faraday rotation is also displayed in Fig.~\ref{fig:PI}. \citet{Vanderwoude+2024} find that the residual off-axis polarization leakage, based on the correction using holographic beam measurements in pilot data, is about 0.5\%. The full-survey POSSUM data is expected to achieve residual off-axis leakage as low as 0.2\% (Gaensler et al. 2024, submitted). We therefore set a cutoff of 8 for signal to noise ratio in $PI$ and 0.5\% for the fractional polarization. The panels in Fig.~\ref{fig:PI} only show pixels where the values are above these cutoffs.

We detect polarized emission from several segments close to the center of the PWN. The fractional polarization ranges from 0.5\% to about 2.5\%. The RMs of the polarized segments range from $-$601~rad~m$^{-2}$ to $-696$~rad~m$^{-2}$, with uncertainties less than about 2~rad~m$^{-2}$. The polarization vectors appear to be coherent in these patches, and the distribution of RMs bears the signature of a toroidal field expected in a PWN \citep[e.g.][]{Kothes2006}. We discuss these in Sect.~\ref{sec:discussion}.

\begin{figure}[ht!]
\centering
\includegraphics[width=0.48\textwidth]{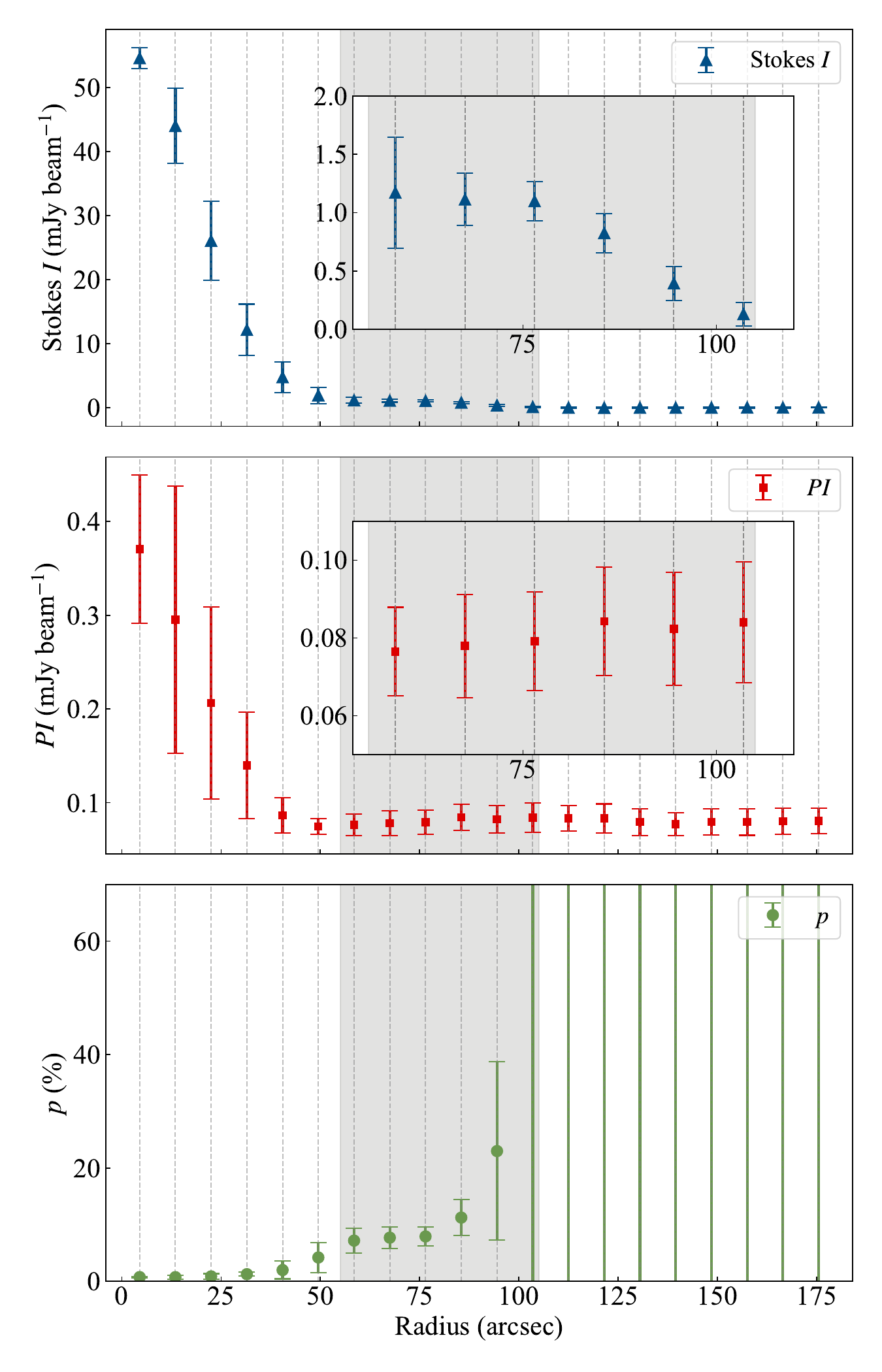}  
\caption{Radial profiles of radio total intensity (Stokes $I$, blue, top panel), polarization intensity ($PI$, red, center panel), and fractional polarization ($p$, green, bottom panel). The inset panel shows the profile across the full radial extent, while the right panel zoom in on y-axis to show the faint structure. The grey region indicates where the shell is located.}
\label{fig:profile_pi}
\end{figure}

For most parts of the remnant, including the shell, the peak polarized intensity image shown in Fig.~\ref{fig:PI} does not show polarized emission above our cutoff values. To look for polarized emission with a higher signal to noise ratio, we take advantage of the shell's symmetry and look at radial profiles of linearly polarized quantities: the polarized intensity ($PI$) and the fractional polarization ($p$). The radial profiles are computed from the $PI$ image (Fig.~\ref{fig:PI}) and the MFS Stokes $I$ image which is convolved to 18$\arcsec$ to match the resolution of the $PI$ image. The centers of the annuli are the same as in Fig.~\ref{fig:profile}. The widths of the annuli are set to 9$\arcsec$ and the radius of the largest annulus is set to 180$\arcsec$. The resulted radial profiles are shown in Fig.~\ref{fig:profile_pi}. 

For the radial profiles of total intensity and polarized intensity, we plot the mean of the points within corresponding annuli in the respective images. The error bars show the corresponding standard deviations of the quantities. For fractional polarization, the radial profile shows the ratio of the $PI$ profile to that of the Stokes $I$ profile. The error bars show the standard deviations (within the respective annuli) in the fractional polarization derived as the ratio of the $PI$ noise image (measured in a signal-free frame with the Faraday depth of 2000~rad~m$^{-2}$) and the Stokes $I$ image.

As can be seen from the radial profile of fractional polarization (see Fig.~\ref{fig:profile_pi}), there is significant polarized emission, particularly in the limb-brightened parts of the SNR shell. This provides additional independent evidence of the emission from the shell being synchrotron in nature. We discuss the details in Sect. \ref{sec:discussion}. 

\section{Discussions} \label{sec:discussion}

\citet{Renaud2010} combined the dispersion measure (DM) of the pulsar PSR J1400$-$6325, the extinction from the HI column density obtained from X-ray spectral fitting, and the relation between spin-down power and X-ray luminosity to derive a distance of 7~kpc for G310.6$-$1.6, which puts the SNR at the Crux-Scutum spiral arm. We adopt this distance for the discussions hereafter.

\subsection{Radio spectro-polarimetry}\label{sec:dp}
 
SNR shells are expected to emit synchrotron emission. While the derived spectral index $\alpha_\text{shell} = -0.7\pm0.3$, discussed in Sect. \ref{sec:radiospectra}, is consistent with synchrotron emission, linearly polarized emission provides a more direct evidence. Detections of polarization are challenging, especially when the emission is weakly polarized, as is the case here. 

From the radial polarization profiles shown in Fig.~\ref{fig:profile_pi}, we find evidence of polarization from the remnant, with $p\sim 5\%$ in the shell region and reaching upwards of $10\%$ near the outer edges of the shell ($\sim 100\arcsec$), beyond which the fractional polarization is dominated by noise. For the central regions of the remnant, the fractional polarization is very low. 

For the whole SNR, the fractional polarization we detect is much less than the theoretical value of $p\sim70$\% expected from synchrotron emission~\citep[e.g.][]{Pacholczyk1970}. The difference between what we observe and this theoretical value can be due to the Stokes $I$ emission having contributions from mechanisms other than synchrotron emission. Low polarized fractions can also be due to depolarization from destructive interference of the polarization vectors either along the line-of-sight (field reversals and/or mixing of Faraday-rotating thermal and synchrotron-emitting non-thermal plasma) or within the beam of the telescope. At low frequencies this depolarization can be enhanced by Faraday rotation effects. At high frequencies, where Faraday rotation is negligible, depolarization is purely due to geometry of the intrinsic magnetic field. 

High frequency (10~GHz) observations of PWNe have detected \(p\)=10\%\ for CTB~87 \citep{Kothes+2020}, and \(p\)=24\%\ for DA~495 \citep{Kothes+2008}. In the case of G310.6$-$1.6, if we assume a similar intrinsic value of \(p\sim20\%\), then our observed polarized fraction of \(p\sim1\%\) at 943~MHz indicates a significant depolarization factor of $\sim0.05$. 
Ascertaining these will require future observations of G310.6$-$1.6 at suitable frequencies and bandwidth.

\subsection{RM}

\begin{figure}[ht!]
\centering
\includegraphics[width=0.45\textwidth]{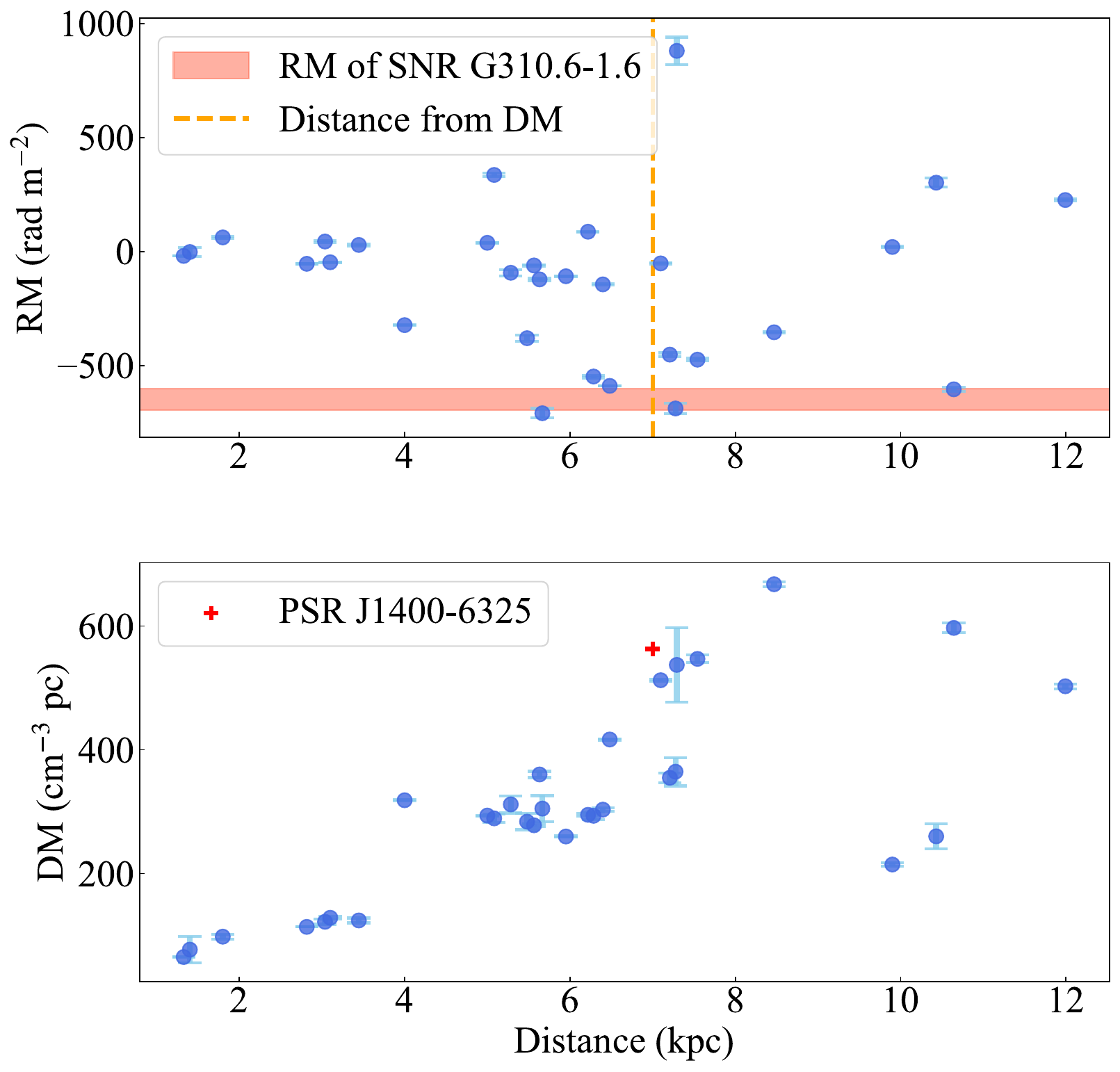}
\caption{RM (top) and DM (bottom) versus distance 
for pulsars within $5\arcdeg$ of PSR J1400$-$6325 which is marked by a cross. The RM value of G310.6$-$1.6 is from  $-$696~rad~m$^{-2}$ to $-$601~rad~m$^{-2}$. The dashed line indicates the distance of 7~kpc.}
\label{fig:RM-DM-dist}
\end{figure}

The RMs of the polarization segments, shown in Fig.~\ref{fig:PI}, include contributions from both the magneto-ionic medium of the PWN itself and also a foreground component that comes from the interstellar medium of the Milky Way along the line-of-sight. In order to analyze the contribution from the PWN, we must first estimate and remove the foreground. We use two methods to estimate the foreground: the first is based on nearby pulsars, and the other on the RM distribution of polarized components detected in the POSSUM survey. We use the term “components” instead of “sources”, as a single background synchrotron source can consist of multiple components, each with its own RM \citep[see][for a detailed explanation]{Vanderwoude+2024}.

We retrieve pulsars within $5\arcdeg$ of PSR J1400$-$6325 from the ATNF pulsar catalog\footnote{https://www.atnf.csiro.au/research/pulsar/psrcat/}~\citep{Manchester2005}, plotting RM and DM against distance\footnote{The best estimate of the pulsar distance using the YMW16~\citep{Yao+2017} DM-based distance as default (kpc)}. in Fig.~\ref{fig:RM-DM-dist}. There is a linear relationship between DM and distance until a distance of about 8~kpc, while the RMs have a large scatter. The median RM for pulsars between 6~kpc and 8~kpc is about $-$450~rad~m$^{-2}$. The three pulsars with angular separations less than $2\degr$ have RMs of $-$474~rad~m$^{-2}$, $-$589~rad~m$^{-2}$, and $-$709~rad~m$^{-2}$, leading to an average value of $-$590~rad~m$^{-2}$ with a scatter of about 120~rad~m$^{-2}$.

\begin{figure}[ht]
    \centering
    \includegraphics[width=0.48\textwidth]{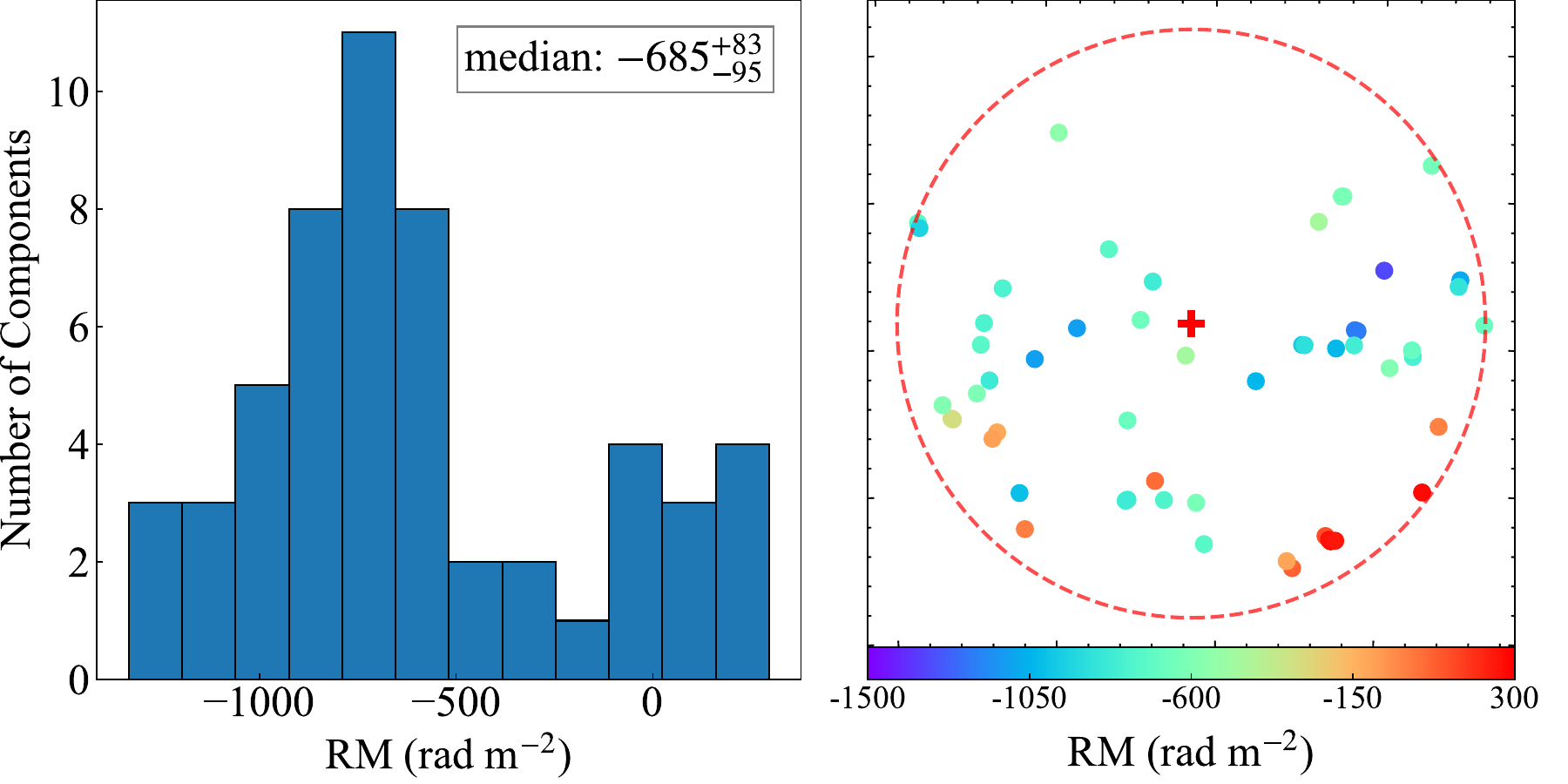}
    \caption{Histogram of RM values of polarized components within $1 \arcdeg$ of the PWN (left) and RM spatial distribution in the J2000 coordinate system (right)}. The red dashed line circle indicates a radius of $1 \arcdeg$, and the red cross indicates the location of SNR G310.6$-$1.6.
    \label{fig:rm_distribution}
\end{figure}

We analyze polarized components located within $1 \arcdeg$ of G310.6$-$1.6. For this analysis, we used the preliminary catalog of polarized components that is available on CASDA. For components that are less than $0.01 \arcdeg$ apart, we treated them as one component and used their average RM values. A total of 54 components were selected and their RM distributions are shown in the histogram (Fig.~\ref{fig:rm_distribution}) along with their locations in the sky. The histogram on the left of Fig. \ref{fig:rm_distribution} indicates that the RM values vary widely, reflecting contributions from both within the polarized components and variations in the foreground RM. To estimate the foreground RM contribution, we calculated the median RM value using bootstrap resampling, obtaining a value of $\rm -685_{-95}^{+83}~rad~m^{-2}$, which agrees with the range of the pulsar RMs. 
This value, derived from more lines of sight closer to the SNR than pulsars, is adopted as the foreground RM for the following analysis. In Fig.~\ref{fig:RM-sub} we show the resulting RM after subtracting the average foreground RM of $\rm -685 ~rad~m^{-2}$.

\subsubsection{RM variation}

\begin{figure}[ht!]
\centering
\includegraphics[width=0.45\textwidth]{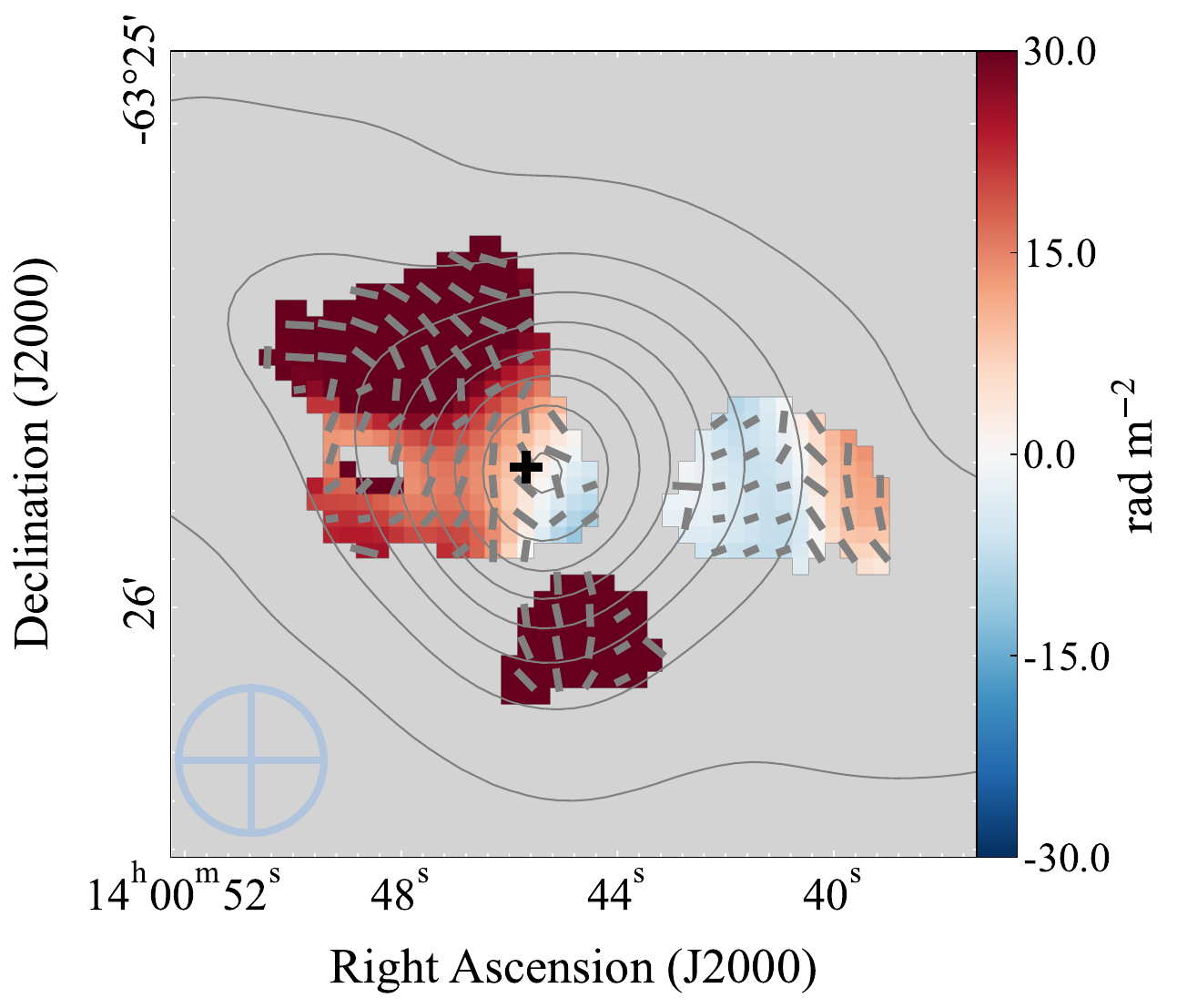}
\caption{RM variation for the central area of PWN after subtracting the foreground RM contribution of $-$685~rad~m$^{-2}$ from the RM map shown in Fig. 6 (right panel)}. The length of the grey bars has been scaled to the square root of the polarized intensity to enhance the visibility of the orientation.
\label{fig:RM-sub}
\end{figure}

The observed RM of the PWN varies between $-$696~rad~m$^{-2}$ and $-$601~rad~m$^{-2}$, and the RM variation is up to about 100~rad~m$^{-2}$. This variation could result from either Galactic RM fluctuations or internal RM within the PWN itself.

To estimate the Galactic RM variation, we use the RM structure function ($D_{\rm{RM}}$) from \citet{Haverkorn2008}. From their Fig.~1, the structure function for the Crux arm, where G310.6$-$1.6 is located, is flat regarding with the angular separation. In this case, $D_{\rm RM}=2\sigma_{\rm RM}^2$, and the RM scattering $\sigma_{\rm RM}$ is about 130~rad~m$^{-2}$. 
This can explain the observed RM variation. However, such a large foreground RM variation would produce a totally distorted polarization angle distribution, which is not the case shown in Fig.~\ref{fig:PI}.

It is more likely that the variation in RM is indicative of internal RMs, meaning that there are thermal electrons inside the PWN. With a foreground RM of $-$685~rad~m$^{-2}$, the internal RM is in the range approximately from $-$85~rad~m$^{-2}$ to $+10$~rad~m$^{-2}$, as shown in Fig.~\ref{fig:RM-sub}. The values of internal RMs depend on the foreground RM which is very uncertain. To produce the observed maximum internal RM difference of about 100~rad~m$^{-2}$, the absolute values of internal RM for parts of the PWN are required to be greater than half of the difference, namely about 50~rad~m$^{-2}$.   

If we assume that relativistic and thermal electrons are uniformly mixed, then the differential Faraday rotation along the line of sight will cause depolarization and the depolarization factor can be calculated as $|\sin(2{\rm \Delta RM}\lambda^2)/(2{\rm \Delta RM}\lambda^2)|$, where $\Delta {\rm RM}$ is the internal RM~\citep{Sokoloff+1998}. The depolarization factor of 0.05 from Sec.~\ref{sec:dp} thus leads to an observed absolute value of the internal RM of 100~rad~m$^{-2}$ or less. Larger absolute values of internal RMs will result in almost total depolarization.

In summary, we conclude that the bulk of the RM contribution comes from the Milky Way foreground. However, the evidence presented here supports the existence of internal RM from the PWN.
We used an absolute value of 50~rad~m$^{-2}$ for the internal RM when estimating the lower limit of the thermal electron density inside the PWN in the analysis below.

\subsection{Magnetic field}

The RM image in Fig.~\ref{fig:RM-sub}~exhibits a positive sign in the northeastern patch and a negative sign in the central and western regions. Note that the positive sign indicates a magnetic field pointing towards us and vice versa for the negative sign. This suggests the existence of a toroidal field, which is very common for PWNe~\citep{Kothes2006}. The polarization angles across these areas remain relatively uniform, implying that the toroidal field is shaping both the RM and the observed polarization features. However, there are further sign changes towards the west, which challenges the interpretation. It could be that there exists an extra poloidal field component~\citep{Reynolds2012}, which disrupts the coherent magnetic field. Higher frequency and higher resolution observations are needed to further investigate these RM patterns and clarify the underlying structures. 

The magnetic field strength can be estimated by assuming the energy equipartition~\citep{Pacholczyk1970} between cosmic ray particles and the magnetic field as 
\begin{equation}
    B=4.5^{2/7}(1+k)^{2/7}c_{12}^{2/7}f^{-2/7}R^{-6/7}L^{2/7}. 
\end{equation}
where $k$ is the ratio of proton energy to electron energy, $f$ is the volume filling factor, $R$ is the radius, $L$ is the radio luminosity and $c_{12}$ is a constant. We used $k=100$ estimated by \citet{Beck2005}, and the frequency range of $10^7$--$10^{10}$~Hz for the integral of luminosity. In this way, the equipartition magnetic field was estimated to be about 120~$\mu$G for the PWN of G310.6$-$1.6. We note that the equipartition condition might not hold for PWNe, but it is still worthwhile to compare the equipartition value with those estimated from other methods, as has been done for PWN G327.1$-$1.1~\citep{Ma2016}.

There are no stringent constraints of the magnetic field from high-energy observations of the PWN. The non-detection of TeV photons put a lower limit of about 6~$\mu$G for the magnetic field~\citep{Renaud2010}. There is a break at about 6~keV~\citep{Renaud2010} or 4.2~keV~\citep{Bamba2022} from X-ray observations. If the break was caused by synchrotron cooling, the magnetic field would be small, about 10~$\mu$G, otherwise G310.6$-$1.6 would be too young~\citep{Renaud2010}. However, the break can be interpreted as the cutoff at high energies for injected electrons~\citep[e.g.][their Fig.~4]{Martin2014}. Interestingly, a model with a magnetic field of up to about 300~$\mu$G can still fit the broad spectrum from radio to TeV~\citep{Martin2014}.

There is a spectral break at the frequency of about $10^{14}$~Hz between radio with $\rm \alpha_{pwn}=-0.4$ and X-ray with $\rm \alpha_{pwn}=-1$ from a photon spectral index of about 2~\citep{Bamba2022}. The break has been well reproduced with a broken power law for the injected electrons~\citep{Tanak2013, Martin2014, Zhu2018}. However, the radio emission extending beyond the X-ray emission~(Fig.~\ref{fig:RC}) also favors the scenario of synchrotron cooling. The magnetic field $B$ in $\mu$G and the age $t_3$ in $10^3$ yr can be related as $B=215t_3^{-2/3}$ \citep{Gaensler2006}. For an age of 2500~yr, the magnetic field would be about 120~$\mu$G, consistent with the estimate from energy equipartition. 

\subsection{Preshock density}

To produce the internal RM of 50~rad~m$^{-2}$ contributed by the PWN, the minimum electron density is required to be about 0.2~cm$^{-3}$, assuming that the magnetic field of 120~$\mu$G is completely parallel to the line of sight and the path length is approximately the radio extent of the PWN from the southeast to the northwest with a radius of about $40\arcsec$~(Fig.~\ref{fig:profile}). The actual electron density has to be much larger because a large fraction of the magnetic field is expected to be perpendicular to the line of sight to generate the observed synchrotron emission. With a much smaller magnetic field of about 10~$\mu$G~\citep[e.g.][]{Renaud2010}, the electron density would be even higher. Furthermore, The PWN evolves within the ejected mass of the progenitor star, where the region with the highest electron density is still at the forward shock of the SNR~\citep{Blondin2001}. The electron density of the shell (preshock density) is not expected to be lower than the electron density derived from the PWN. 
 
The radio luminosity of the SNR shell is about $2\times10^{21}$~erg~s$^{-1}$~Hz$^{-1}$ at 1~GHz with a spectral index of $\rm \alpha_\text{shell} = -0.7\pm0.3$. To produce such a low luminosity, either the supernova explosion energy, $E_{\rm sn}$, or the preshock density $n_0$ is low. According to \citet{Berezhko2004} and their Fig.~4, to derive the observed luminosity for the SNR shell with a diameter of $2R_{\rm snr}\approx6.8$~pc, $E_{\rm sn}$ is around $1.6\times10^{48}$~erg with $n_0=1$~cm$^{-3}$, and $E_{\rm sn}$ is around $5\times10^{49}$~erg with $n_0=0.1$~cm$^{-3}$. Note that the simulations by \citet{Berezhko2004} used an ejected mass $M_{\rm ej}=1.4 M_\sun$. If the ejected mass is increased to a typical value of $10 M_\sun$, the explosion energy is correspondingly increased to about $1.3\times10^{50}$~erg to produce the same radio luminosity, according to \citet{Berezhko2004}.

We take $E_{\rm sn} = 1.3 \times 10^{50}~{\rm erg}$, $n_0 = 0.1~{\rm cm}^{-3}$, and $M_{\rm ej} = 10~M_\odot$ for the analysis below. These values are also consistent with the correlation between $E_{\rm sn}$ and $M_{\rm ej}$ derived by \citet{Pejcha+2015}. 
We note that the preshock density used in this study is larger than the value of 0.01 cm$^{-3}$ given by \citet{Reynolds2019} using X-ray analysis, but consistent with our updated X-ray analysis (see Sect. \ref{sec:xrayspectra}).

\subsection{Age}

Suppose a spherical explosion with a density of $n_0=0.1$~cm$^{-3}$, the mass swept up by the supernova remnant (SNR) shell is about $0.6 M_\sun$, which is much smaller than the ejected mass. This means that the SNR shell is likely still in the free expansion phase. The expansion speed can be estimated as $\sqrt{2E_{\rm sn}/M_{\rm ej}}$, which is about 1200~km~s$^{-1}$, implying an age of about 2770~yr. This roughly agrees with the age of 2500~yr estimated from synchrotron cooling. 

The ratio of the PWN radius to the radius of the SNR shell ($\mathcal{R}$) indicates the evolution of the object~\citep[e.g.][]{VanDerSwaluw2001}. $\mathcal{R}$ can be better determined from the radio image than from the X-ray image because the life time of the radio-emitting electrons is much longer than that of the X-ray-emitting electrons.
As can be seen in Fig.~\ref{fig:RC}, $\mathcal{R}$ is about 0.4 from the southeast to the northwest and is close to 0.9 from the southwest to the northeast. 

According to \citet{VanDerSwaluw2001}, the ratio can be represented as $\mathcal{R}(t)\approx\eta_3(t)(E_{\rm sd}/E_{\rm sn})^{1/3}$, where $\eta_3(t)$ was introduced as additional dimensionless parameter and is between 1 and 3 for the first several thousand years and almost constant close to 1 afterward, and $E_{\rm sd}$ is the spin-down energy driving the PWN. For an age $t=2500$~yr, $E_{\rm sd}\approx \dot{E} t$ is about $4.02\times10^{48}$~erg, where $\dot{E}$ is the spin-down luminosity of the pulsar. These values lead to $\mathcal{R}(t)\approx0.3\eta_3(t)$. With $\eta_3$ of about 3, the observed $\mathcal{R}$ from the radio image can be explained. 

The radius of the PWN can be estimated with $\dot{E}$, $E_{\rm sn}$, $M_{\rm ej}$, and $t$~\citep{Gaensler2006}, which is about 1.6~pc with $t=2500$~yr. This is consistent with the PWN radius of about $40\arcsec$ or 1.4~pc from the southeast to the northwest.

In addition, the steep radio spectral index of $\rm \alpha_\text{shell} = -0.7\pm0.3$ is typical for a young SNR still in the free expansion phase \citep{Ranasinghe2023}. From the X-ray expansion study, a lower limit of 2500~yr for the age was derived by \citet{Reynolds2019}. These corroborate our estimate, and we use the age of 2500~yr for the analysis below. 

\subsection{Origin of the SNR shell} 

\citet{Reynolds2019} presented several scenarios to interpret the SNR shell, such as the blast wave, reverse shock, pulsar-fed emission, and low-energy supernovae, but none are conclusive. Based on our analyses above, we proposed an $E_{\rm sn}$ of $1.3 \times 10^{50}~{\rm erg}$, much larger than the upper limit $E_{\rm sn}$ of $3 \times 10^{47}~{\rm erg}$ by \citet{Reynolds2019}. We thus only discuss the first three scenarios below.

\subsubsection{Blast wave} 
In this scenario, both the X-ray and radio emissions are from the forward shock. The X-ray emission is from the TeV electrons, and thus traces the front of the forward shock where the acceleration efficiency is high. Because of synchrotron cooling, only the recently accelerated electrons produce the observed X-rays. In contrast, radio emission is from lower energy electrons and cooling can usually be neglected. Therefore, it is expected that the X-ray emission is outside of the radio emission, as has been shown by theoretical calculations and simulations~\citep{Reynolds1981,Cassam2005}. This has also been observed for SNRs with synchrotron X-ray emission, such as SNR G347.3$-$0.5~\citep[][their Fig.~9]{Lazendic2004}, and SNR G32.4+0.1~\citep[][their Fig.~4]{Reynolds2024}. However, this is in contradiction to the observations shown in Figs.~\ref{fig:RC} and \ref{fig:profile}, where the radio emission is clearly outside the X-ray emission. Therefore, the blast wave scenario is not favored. 

Another assumption is that the mismatch between the X-ray and radio emission from the forward shock are caused by expansion. If we consider the mismatch of the shell starting from 2016 and evolving to 2023, the expansion speed would reach 1.28$\arcsec$ yr$^{-1}$, or approximately 40,000 km s$^{-1}$ at a distance of 7~kpc, corresponding to an expansion rate of 1.9\%~yr$^{-1}$. The results conflict with the upper limit of the shock expansion velocity of just 1000 km s$^{-1}$ given by \citet{Reynolds2019}. Longer-term observations in the future, especially high-resolution X-ray observations, will help address this issue by analyzing the expansion velocity of the blast wave.

\subsubsection{Reverse shock} 
The inward-facing reverse shock can also accelerate particles based on the theoretical modeling of~\citet{Ellison2005}. Observational evidence of electrons accelerated to TeV by reverse shock has been found for SNRs Cas~A~\citep{Sato2018} and RCW 86~\citep{Rho2002}. The upstream magnetic field for the reverse shock is much weaker than the ambient magnetic field facing the forward shock. Unlike forward shocks, only some reverse shocks can amplify this weak magnetic field to the extent required for high acceleration efficiency. Therefore, it is expected that the majority of the hard X-ray emission is from forward shock and only a small fraction is from reverse shock, as seen from both Cas~A and RCW~86 and from simulations~\citep{Zirakashvili10}. 

The inward motion of the reverse shock begins at\citep{Blondin2001}:
\begin{equation}
    t\approx 120 (M_{\rm ej}/M_\sun)^{5/6}n_0^{-1/3}(E_{\rm sn}/10^{51}\,{\rm erg})^{-1/2}\rm{yr.}
\end{equation}

It leads to about 4800~yr for G310.6$-$1.6. This implies that the acceleration by the reverse shock has not yet begun. Even if reverse shock acceleration occurred, it is puzzling that the whole X-ray shell is inside the radio shell~(Fig.~\ref{fig:RC}), totally different from Cas~A and RCW~86. Therefore, it is certainly imaginable for the X-ray shell to be produced by reverse shock, but challenges remain to make it work practically.

\subsubsection{Pulsar-fed emission}
The shape of the PWN depends mainly on the interaction between the pulsar kick velocity, the external density gradient, and magnetic field~\citep{Blondin2001,Vander+2003,Swaluw2004}. The simulations by \citet{Kolb2017} showed that an external gradient could produce an asymmetric PWN and a circular shell with a pulsar in the center, as in the case of G310.6$-$1.6. 

In the pulsar-fed emission scenario, the high-energy electrons producing the radio and X-ray emission of the SNR shell are both from the pulsar, and the radio peak is expected to be outside the X-ray peak taking into account the cooling. To reach the radio peak at about $75\arcsec$~(Fig.~\ref{fig:profile}) with an age of about 2500~yr, the diffusion coefficient is about $3.9\times10^{26}$~cm$^2$~s$^{-1}$, reasonable for particles transporting in PWN~\citep[e.g.][]{Tang2012}. The diffusion coefficient is usually energy independent~\citep{Porth2016}. Therefore, to reach the X-ray peak at about $66\arcsec$~(Fig.~\ref{fig:profile}),  the cooling age of the electrons responsible for the X-ray emission shall be around 1900~yr. The cooling age can be represented with magnetic field $B$ (in $\mu$G) and the energy of the emitting photons $E_{\rm ph}$ (in keV) as $55.2~(B/100)^{-3/2}E^{-1/2}_{\rm ph}$~yr~\citep{Olmi2023}. For $E_{\rm ph}=2$~keV, the magnetic field is about 7.5~$\mu$G, which is also reasonable.  

For the pulsar-fed emission scenario to work, it would require the supernovae to explode inside a spherical cavity. Neutral hydrogen (HI) observations are typically used to detect cavities. However, the small angular size of G310.6$-$1.6 requires sufficiently high-resolution HI data for detection, which are unfortunately not available yet. We still examined low-resolution datasets from the HI 4$\pi$ survey (HI4PI, \citealt{HI4PI}) survey but could not find any cavities associated with G310.6$-$1.6.  

We have also searched infrared data for possible evidence of cavities, such as the \textit{Wide-field Infrared Survey Explorer} (WISE, \citealt{Wright2010}) data at 22 $\mu$m and 12 $\mu$m bands, and the AKARI far-infrared all-sky survey~\citep{Doi2015} at 60 $\mu$m, 90 $\mu$m, 140 $\mu$m, and 160 $\mu$m bands. There is no evidence of the cavity that can be identified. 

From the radio morphology, the PWN appears to connect with the shell in the northeast and southwest, which leads to an intriguing possibility: High-energy electrons could be tunneled into the shell from the central bar~(Fig.~\ref{fig:RC}). 

Radio observations at higher frequencies are required to establish a spectral index map of G310.6$-$1.6 to investigate the connection between the PWN and the SNR shell, which is planned and will be presented in a future paper.

\subsection{Comparison with other SNRs}

Among Galactic SNRs, G310.6$-$1.6 is unique and has one of the lowest surface brightnesses. It displays two outstanding characteristics: a central PWN with a circular shell putting it in the group of symmetrical composite SNR; a nonthermal X-ray shell putting it into the group of XSSNR. 
In the context of each group, it possesses an exclusive nature.

\subsubsection{Symmetrical composite SNR}
Noticeable SNRs in this group are G11.2$-$0.3, G292.0$+$1.8, G21.5$-$0.9, which have been discovered to manifest shell structure in X-ray. 
G11.2$-$0.3 is  one of the youngest core-collapse SNRs in the Milky Way with an age of 1400--2400 yr~\citep{Borkowski2016}. 
The radio emission from the SNR shell is brighter than that from the PWN, opposite to G310.6$-$1.6. 
There is also hard X-ray emission in 5--8 keV from the shell, but its spectrum can be well fit with a combination of \textit{vpshock} and \textit{srcut} modes in XSPEC, and the thermal emission is dominant. 

G292.0$+$1.8 is also a core-collpase SNR with an age of around 2500~yr~\citep{Gaensler2003}. It is an oxygen-rich SNR as can be seen from X-ray observations. The ratio of the radius of the PWN to the radius of the SNR is also large in both radio and X-ray, similiar to G310.6$-$1.6. The SNR is surrounded by a soft X-ray shell~\citep{Bhalerao2019}.

G21.5$-$0.9 is also a young SNR with an age in the range 200--1000~yr~\citep{Bocchino2005}. Similiarly to G310.6$-$1.6, G21.5$-$0.9 has a limb-brightened X-ray shell that is non-thermal~\citep{Bocchino2005,Guest2019}. However, no radio shell has been detected so far~\citep{Bietenholz2008, Bietenholz2011}.

Recently, another young SNR, G329.9$-$0.5, has been discovered~\citep{Smeaton+2024}. Although it has a similar morphology as G310.6$-$1.6 in radio,  the correlation between X-ray and radio is still inconclusive.

\subsubsection{XSSNR}

XSSNRs are the best candidates for studying particle acceleration mechanisms in SNRs. However, as of now, only eight such remnants have been identified in our Galaxy including G310.6$-$1.6~\citep{Reynolds2024}. Given the limited number of samples, each new discovery of XSSNR could provide crucial insights. Except for G310.6$-$1.6, all other XSSNRs are shell-type SNRs. Some of them may also be associated with a PWN, but such associations await confirmation.

SNRs in this group, such as G1.9$+$0.3 and SN~1006 have well defined radio and X-ray shells. Their X-ray shells are bilateral in shape, and the hard X-ray shell is outside the radio shell~\citep{Reynolds2008, Cassam2008}. In contrast, G310.6$-$1.6 has complete circular radio and X-ray shells, and the radio shell is outside the X-ray shell. For G1.9$+$0.3, the X-ray shell is brightest toward the east and west, whereas the radio shell is brightest toward the north. The mismatch has been attributed to the interaction with a molecular cloud~\citep{Enokiya2023} or the asymmetric circumstellar medium~\citep{Villagran2024}. For SN~1006, the radio and X-ray shells with enhancement in the northeast and southwest limbs are most likely due to a polar cap geometry \citep{Katsuda2017}.

\section{conclusions}\label{sec:conclusions}

We presented images of total intensity and polarized intensity of G310.6$-$1.6 at 943~MHz from the ASKAP EMU and POSSUM surveys. 

The weak radio shell with a surface brightness of about $7\times10^{-22}$~W~Hz$^{-1}$~sr$^{-1}$~m$^{-2}$ was detected here at 943~MHz. The in-band radio spectral index is $-0.7\pm0.3$. 

We observed strong radio emission from the central PWN, and obtained an in-band spectral index of $-0.4\pm0.1$. Compared with X-ray observations, a spectral break at a frequency of about $10^{14}$~Hz is expected, which is probably caused by synchrotron cooling. The energy equipartition magnetic field was estimated to be about 120~$\mu$G. We also detected polarized emission and obtained an RM between $-$696~rad~m$^{-2}$ and $-$601~rad~m$^{-2}$. We have also estimated RM contribution from the Milky Way as 
$\rm -685_{-95}^{+83}~\mathrm{rad~m^{-2}}$. The variation of RM and depolarization imply internal RM with a typical value of about 50~rad~m$^{-2}$. The variation in the sign of the RM suggests a possible toroidal field. 

We suggested an age of about 2500~yr, a supernova explosion energy of $1.3\times10^{50}$~erg, an ejected mass of $10\,M_\sun$, and a preshock density of $0.1$~cm$^{-3}$ for G310.6$-$1.6. Subsequently, the low radio luminosity of the shell, the spectral break at $10^{14}$~Hz for the PWN, the internal RM of the PWN, and the large ratio between the PWN radius and the SNR radius in radio can be explained.   

The radio circular shell outside the hard X-ray circular shell is very unique among SNRs and yet to be understood. Possible scenarios include blast wave, reverse shock, and pulsar-fed emission. However, none of them are able to explain the offsets between the two shells. Further exploration through radio and high-energy observations will be crucial to fully elucidate the origin and evolutionary state of SNR G310.6$-$1.6.

Our results demonstrate the potential for discovering new objects of small angular size and low surface brightness for the ongoing ASKAP EMU and POSSUM surveys.

\section*{Acknowledgments}
This research has been supported by the National SKA Program of China (2022SKA0120101). 
P.Z thanks the support from NSFC grant No.12273010. M.D.F and S.L. acknowledge Australian Research Council (ARC) funding through grant DP200100784.
W.-H. Jing is supported by the Scientific Research Fund Project of Yunnan Education Department (Project ID: 2025Y0015) and Scientific Research and Innovation Project of Postgraduate Students in the Academic Degree of Yunnan University (Project ID: KC-24248558). We thank Dr. Wolfgang Reich, and Dr. Michał J. Michałowski for careful reading of the manuscript and providing valuable comments. 

This scientific work uses data obtained from Inyarrimanha Ilgari Bundara / the Murchison Radio-astronomy Observatory. We acknowledge the Wajarri Yamaji People as the Traditional Owners and native title holders of the Observatory site. The Australian SKA Pathfinder is part of the Australia Telescope National Facility \footnote{https://ror.org/05qajvd42} which is managed by CSIRO. Operation of ASKAP is funded by the Australian Government with support from the National Collaborative Research Infrastructure Strategy. ASKAP uses the resources of the Pawsey Supercomputing Centre. Establishment of ASKAP, the Murchison Radio-astronomy Observatory and the Pawsey Supercomputing Centre are initiatives of the Australian Government, with support from the Government of Western Australia and the Science and Industry Endowment Fund. The POSSUM project \footnote{https://possum-survey.org} has been made possible through funding from the Australian Research Council, the Natural Sciences and Engineering Research Council of Canada, the Canada Research Chairs Program, and the Canada Foundation for Innovation.
This research has made use of data obtained from the Chandra Data Archive provided by the Chandra X-ray Center (CXC).

\vspace{5mm}
\facilities{ASKAP, CXO}

\software{ASKAPsoft \citep{Guzman2019},
          Astronomy-oriented Python packages: astropy \citep{astropy:2013,astropy:2018,astropy:2022}, 
          pyregion,
          General-purpose Python packages: matplotlib \citep{Hunter:2007}, 
          numpy \citep{numpy}
}

\bibliography{main}
\bibliographystyle{aasjournal}

\end{document}